\documentstyle[preprint,prb,aps,graphicx,epsf]{revtex}

\begin{document}
\begin{center}  
{\large \bf The Effect of Structural Distortions on the Electronic 
Structure of Carbon Nanotubes}\\
\vskip0.5cm
{\bf Alain Rochefort\footnote{e-mail:
rochefor@cerca.umontreal.ca}$^{,{\dagger}}$ Dennis~R.~Salahub\footnote{e-mail:
Dennis.Salahub@umontreal.ca}$^{,{\dagger},{\ddagger}}$ and 
Phaedon~Avouris\footnote{e-mail: avouris@us.ibm.com}$^{,{\S}}$}\\ 
{\small \it $^{\dagger}$ Centre de Recherche en Calcul Appliqu\'e (CERCA), 5160 boul. 
D\'ecarie,\\
bureau 400, Montr\'eal, (Qu\'ebec) Canada H3X 2H9\\ 
$^{\ddagger}$ D\'epartement de Chimie, Universit\'e de Montr\'eal, C.P. 6128,\\ 
Succ. Centre-Ville, Montr\'eal, (Qu\'e) Canada H3C 3J7\\ 
$^{{\S}}$ IBM Research Division, T.J. Watson Research Center, P.O. Box 218,\\ 
Yorktown Heights, NY 10598, USA}
\vskip0.5cm

{\bf Abstract}\\
\end{center}

\noindent

{We calculated the effects of structural distortions on the electronic
structure of carbon nanotubes. The main effect of nanotube bending is an
increased mixing of $\sigma$ and $\pi$-states.  This mixing leads to an
enhanced density-of-states in the valence band near the Fermi energy. While
in a straight tube the states accessible for electrical conduction are
essentially pure C($2p_{\pi}$)-states, they acquire significant
C($2sp_{\sigma}$) character upon bending.  Bending also leads to a charge
polarization of the C-C bonds in the deformed region reminiscent of interface
dipole formation. Scattering of conduction electrons at the distorted regions
may lead to electron localization.}

\section{Introduction}

Carbon nanotubes are an interesting class of nanostructures which can be
thought of as arising from the folding of a graphene sheet.  Depending on the
width of the graphene sheet and the way it is folded a variety of different
nanotube structures are formed.  The nanotubes are usually described using
the chiral vector:  $C_{h} = n\vec{a}_{1} + m\vec{a}_{2}$, where
$\vec{a}_{1}$ and $\vec{a}_{2}$ are unit vectors of the hexagonal honeycomb
lattice and $n, m$ are integers, and a chiral angle $\theta$ which is the
angle of the chiral vector with respect to the zigzag direction of the
graphene sheet \cite{saito,dresselhaus}. The one-dimensional electronic
structure of a nanotube with indices $(n, m)$ can be predicted on the basis
of the two-dimensional electronic structure of graphite. Armchair $(n, n)$
nanotubes have a band degeneracy between the highest $\pi$-valence band and
the lowest $\pi^*$-conduction band at $k = \pm (2/3)(\pi /a_{0})$, where
these bands meet the Fermi level, and should show metallic behavior. Among
the other $(n, m)$ nanotubes, the ones with $n - m = 3i$ (where $i$ is an
integer) should also be metallic, while the rest should have a band-gap and,
therefore, be semiconductors.\cite{saito,dresselhaus} Recent STM spectroscopy
experiments verified these predictions \cite{wildoer,odom}. Their electrical
properties coupled with the superb mechanical strength of the nanotubes
\cite{treacy,wong} makes these materials promising candidates for use as ideal
one-dimensional (1-D) conductors, if ballistic transport can be achieved or,
in the case of semiconducting tubes, as key building elements of novel
nanoelectronic devices \cite{tans,martel}.\\

The above discussion of the properties of carbon nanotubes is based on the
assumption that they have the perfect symmetry expected for a free nanotube.
Nanotubes, however, are usually studied and utilized while being supported on
a solid substrate.  Studies of supported nanotubes using atomic force
microsopy \cite{hertel1}  and molecular mechanics calculations \cite{hertel2}
show that supported nanotubes can undergo significant axial and radial
distortions.  The nanotubes tend to bend so as to conform to the morphology
of the substrate and thus optimize the van der Waals adhesion forces.
Similarly, the side of the nanotube in contact with the surface may flatten
so as to optimize the contact area. The extent of the above distortions
depends on the balance between the increased adhesion and the rise in strain
energy produced by the distortions. These structural distortions can, in
turn, modify the nanotube's electronic structure and electrical transport
properties.  For example, a reduction in symmetry may result in the lifting
of degeneracies. An increased curvature can enhance $\sigma$-$\pi$ mixing and
rehybridization, while bond strain can modify the band-gap of semiconducting
tubes.  These effects may produce local barriers at the distortion regions
and affect electrical transport.  Previous theoretical work on this problem
\cite{kane}  focused only on the $\pi$-electrons, ignoring the
$\sigma$-electrons and the possibility of $\pi$-$\sigma$ mixing, and
concluded that the effects of bending distortions on electronic structure are
insignificant.  Experimentally, bends in quantum wires produced by patterning
of two-dimensional electron-gas systems have been shown to affect transport
and lead to interesting interference phenomena \cite{wu,weisshaar}.  It is thus likely
that similar effects may be induced by bends in nanotubes and be observable
in low temperature transport experiments.\\

Here we explore the effects of structural distortions on the electronic
structure of nanotubes by performing Extended H\"uckel calculations  on
straight and bent armchair $(6, 6)$ single-wall nanotubes. Important changes
in the local density of states (LDOS) of $\sigma$ and $\pi$-electrons and
increased $\pi$-$\sigma$ mixing are observed to develop as the nanotube is
distorted. These changes become stronger with increasing bending angle.  In
addition, a charge polarization of the C-C bonds in the distorted regions is
observed. The electronic structure changes are expected to have important
implications for the low temperature electrical transport properties of the
nanotubes, and should also affect locally their chemical reactivity.

\section{Computational Details}

Electronic structure calculations were performed on a 972 atom cluster model
(948 C and 24 H that saturate the dangling bonds at the ends) of a
$(6,6)$  nanotube using the extended H\"uckel method\cite{yaehmop}. The
parameters used for carbon and hydrogen were: C$_{2s}$ ($H_{ii}$ = -21.4 eV,
$\zeta$ = 1.625), C$_{2p}$ ($H_{ii}$ = -11.4 eV, $\zeta$ = 1.625), and
H$_{1s}$ ($H_{ii}$ = -13.6 eV, $\zeta$ = 1.3), where $H_{ii}$ and $\zeta$ are
the orbital energy and Slater exponents, respectively.\\

In the undistorted nanotube, the C-C and C-H bond lengths were fixed at the
values obtained for bulk graphite \cite{graphite} that are 1.42 and 1.09
{\AA}, respectively. We have built models with bending angles of 30, 45 and
60$^{\circ}$ using a constant length segment in the middle of the tube to
introduce the distortion. Prior to the electronic structure calculations, the
geometry of the bent nanotubes was optimized with the molecular mechanics
program TINKER \cite{tinker} employing the MM3 force field \cite{mm3}. In the
structural relaxation and optimization process, the first 5 sections (a
section is defined as a single circular plane of carbon atoms that are packed
along the length of the nanotube) at the two ends of the nanotube were kept
fixed in order to maintain the bending angle. In the molecular mechanics
calculations, we used the MM3 alkene  parameters except for the bond length
parameter which we modified from that of an alkene (1.33 {\AA}) to that of
bulk graphite (1.42 {\AA}). Density of states (DOS) plots were generated by
convoluting the computed electronic structure with a 50:50 combination of
Gaussian and Lorentzian functions. In order to analyze the origin of the band
states, we performed a series of projections of the DOS where each molecular
orbital was weighted by the contribution obtained from a Mulliken analysis of
specific carbon atoms.

\section{Results and Discussion}

Figure 1 shows the computed electronic structure of a perfect $(6,6)$
armchair nanotube model. This model contains 948 carbon atoms distributed in
79 circular sections. The overall DOS spectrum shows high binding energy (BE)
states extending from -22 to -12 eV that are mainly $\sigma$-bond states
involving C($2s$) orbitals. The $\sigma$-bond states associated with C($2p$)
orbitals lie at lower BE between -12 and 0 eV with respect to the Fermi
energy ($E_{F}$). The fine structure visible in the high BE region involves
van Hove singularities characteristic of one-dimensional
systems \cite{ashcroft}. Each of these peaks is characterized by a specific
number of nodes in the wavefunction along the circumference of a single
nanotube section, while the 1/$\sqrt{E}$ tail reflects the free electron
character along the tube axis. The more intense DOS bands between -6 and 0 eV
are due to overlapping $\pi$- and $\sigma$-bond states, with the $\pi$-states
centered at a slightly lower binding energy.  The set of DOS bands between 0
and 8 eV are essentially ${\pi}^{*}$-states.  The ${\sigma}^{*}$-states (not
shown) extend above 10 eV. The valence and conduction band states near the
``gap'' edges have $\pi$-character; electric transport involves
$\pi$ electrons. A more detailed view of the DOS in the region around the
Fermi level is shown in the inset of Figure 1.\\

The local density of states (LDOS) profiles presented in Figure 1 where the
projection is summed over the contribution of the twelve carbon atoms
contained in a particular section of the tube shows that the electronic
structure along the tube is quite uniform. The similarity of the LDOS
profiles indicates that the wavefunction of this finite size tube is
delocalized over its entire length.\\

Figure 2 shows the model structures used to evaluate the effects of bending
on the electronic properties of carbon nanotubes. The following approach is
used to generate the 30$^{\circ}$, 45$^{\circ}$ and 60$^{\circ}$ bend models:
a section of constant length in the middle of the tube is bent and the atomic
structure of the nanotube is then optimized with molecular mechanics. As the
bending angle increases, the deformation of the atomic structure of the tube
increases, particularly in the central region of the tube. The 30$^{\circ}$
bend induces a compression of the C-C bonds in the inner side of the tube
while the bonds on the outer side are stretched. The general tubular shape,
however, remains essentially intact in both straight and bent regions. After
a 45$^{\circ}$ bend, the increased compression of C-C bonds in the center of
the nanotube leads to a flattening of its cross-section. Further bending
increases this flattening to the point where the force between opposite
nanotube walls becomes high enough to induce the formation of a kink as in
the 60$^{\circ}$ model of Figure 2. The front views of the central sections
of the bent tube models clearly show the drastic structural deformations that
occur as the bending angle is increased from 45$^{\circ}$ to 60$^{\circ}$.\\

The changes in the electronic structure (from -25 to 10 eV) induced by the
bending of the armchair $(6,6)$ nanotube appear weakly in the total density
of states spectrum.  This is because the deformed region is relatively small
compared to the total length of the nanotube. On the other hand, the LDOS
along the nanotube length. i.e. the LDOS generated by adding the
contributions of the twelve carbon atoms contained in a particular section of
the nanotube, provides a much clearer view of the changes brought about by
bending. As already discussed, in a straight (0$^{\circ}$) nanotube there is
no obvious variation of the LDOS along its length (Figure 1). Upon bending,
the most obvious change is a broadening of the fine structure of the
overlapping $\sigma-\pi$ states near $E_{F}$, which becomes particularly
large in the 60$^{\circ}$ bent example. Far from the distorted region, the
LDOS profiles are very similar (in terms of band position and structure) in
all bent nanotube models. In the distorted segment of the the tube (i.e. near
the central section), the fine structure of the ${\sigma}-{\pi}$ band
disappears. The $\sigma$-bond states at higher binding energy arising from a
combination of C$(2s)$ orbitals are similarly altered.  In general,
$\sigma$-bonds in the bent region appear to be the most perturbed by the
deformation, while $\pi$-bonds are affected to a lesser extent.\\

Another interesting result of the deformation involves the distribution of
net charge along the length of the nanotube shown in Figure 3. Upon
increasing the bending angle from 0$^{\circ}$ to 60$^{\circ}$, the net charge
fluctuation rises.  Furthermore, the position in the bent region where the
largest fluctuations are observed changes with bending angle. A smooth
30$^{\circ}$ bending leads to a broad distribution of small charges among all
carbon atoms in the bent region. A bending of 45$^{\circ}$ gives a larger
charge fluctuation in the innermost deformed region. Finally, the kink in the
60$^{\circ}$ bent nanotube has a particularly dramatic effect on both the
magnitude and the spatial extent of the net charge distribution leading to
large charge fluctuations displaced toward the central edges of the kinked
region.  In general, the largest fluctuations are found where the changes in
the nanotube geometry ($R_{C-C}$) are most pronounced. This localization of
charge fluctuations is directly related to the changes observed in the
$\sigma$-wavefunction discussed above; the larger fluctuations arise from an
enhanced $\sigma$-orbital contribution or, in other words, from an increased
$sp^{2}-sp^{3}$ rehybridization.\\

To investigate how the electrical transport properties of the nanotube may be
affected by bending, we focus on the $\pi$-states near $E_F$.  The upper part
of Figure 4 shows an expanded view of the DOS in this energy range. Bending
is found to induce an increased DOS near $E_{F}$ (at about -1 eV BE).
Furthermore, the DOS at $E_{F}$  remains finite in all bent nanotubes
studied.  More informative are LDOS curves along the nanotube length. In the
case of a straight tube, the lack of change in LDOS along the tube indicates
that ${\pi}$ and ${\pi}^*$ states near $E_{F}$ are not confined in a specific
area but are delocalized along its length. Upon bending, small changes in the
electronic structure are observed at the tube boundaries. These changes
result from the relaxation of the atomic structure of the entire nanotube
upon bending.  The largest changes are found, however, at the centers of the
nanotubes (section 40), in particular in the kinked tube (60$^{\circ}$
angle).  Figure 4 shows an overall increase in the LDOS near $E_{F}$ in the
distorted regions of the nanotubes. Analysis of the wavefunction indicates
that the induced density contains an increased contribution of
C($2sp_{\sigma}$) states.  The shift of the valence band edge to higher
energy and the increased C($2sp_{\sigma}$) contribution can both be explained
by a bending-induced $\pi$-$\sigma$ hybridization.  Such a rehybridization
already exists to some extent in the straight tube due to its curvature
\cite{blase}, but becomes enhanced by further bending.\\

In conclusion, we have shown that the local electronic structure of carbon
nanotubes is modified as a result of structural distortions.  Such
distortions can be the result of the interaction of the nanotubes with the
topography of the substrate they are placed on \cite{hertel2}, with the metal
electrodes used to monitor their electrical properties \cite{bezryadin}, or
as a result of controlled manipulation \cite{hertel1}. Two types of
perturbations of the electronic structure as a result of bending are
predicted by our calculations: First, a modification of the LDOS in the
deformed region of the nanotube leading to an increased DOS near $E_F$ and a
shift of the valence band edge to a lower binding energy.  These effects are
due to increased $\pi$-$\sigma$ hybridization. Discussion of electrical
transport in distorted tubes must take into account this hybridization.
Secondly, we observe a charge polarization of the C-C bonds in the deformed
region. The magnitude of this polarization increases with the bending angle
and is particularly severe upon kink formation.  The above changes in the
local electronic structure are expected to scatter the conduction electrons
at the deformed regions leading to carrier localization, especially at low
temperatures.  Truly ballistic transport in nanotubes may require perfectly
straight tubes \cite{frank}. This conclusion is supported by recent
experiments by Bezryadin {\it et al.} \cite{bezryadin} who examined the
electrical behavior of a chiral single-wall nanotube draped over a set of
several Pt electrodes.  Their studies indicated that electrically the
nanotube was broken into a series of isolated islands as a result of barriers
generated by the bending of the tube over the raised electrodes.\\

{\Large \bf Acknowledgements}\\

We would like to thank T. Hertel for many helpful discussions.\\

\newpage

\begin{figure}
\caption{Total (DOS) and local density of states (LDOS) diagrams of a 
straight armchair $(6,6)$ nanotube (resolution = 0.2 eV). The indices
in the LDOS diagram give the relative position of the carbon atoms in 
the nanotube structure (1: boundary, 40: middle of the nanotube)
The inset gives an expanded view of the DOS near the Fermi level (E=0 eV). 
The zero of the DOS scale is indicated by the
horizontal line, and the energy resolution is 0.05 eV.} 
\label{Figure 1.}
\end{figure}

\begin{figure}
\caption{Structures of bent (6,6) nanotubes optimized using molecular 
mechanics. The bending angles are (from left) 0$^{\circ}$, 30$^{\circ}$, 
45$^{\circ}$ and 60$^{\circ}$. A front view of the most central sections 
of bent models is also shown.}
\label{Figure 2.}
\end{figure}

\begin{figure}
\caption{Variation of the net charge distribution along the length of bent 
nanotubes. Atom indices give the relative position of the carbon atoms in 
the nanotube structure (a total of 948 carbon atoms distributed in 79 
sections of 12 carbon atoms were used)}
\label{Figure 3.}
\end{figure}

\begin{figure}
\caption{Variation of DOS and LDOS near the Fermi level for several bent 
$(6,6)$ armchair nanotubes. (resolution = 0.05 eV). Indices give the 
relative position of the section in the nanotube structure (1: boundary, 
40: middle of the nanotube)}
\label{Figure 4.}
\end{figure}

\clearpage
\newpage

\begin{figure}[p] 
\begin{center}
\includegraphics[width=15cm]{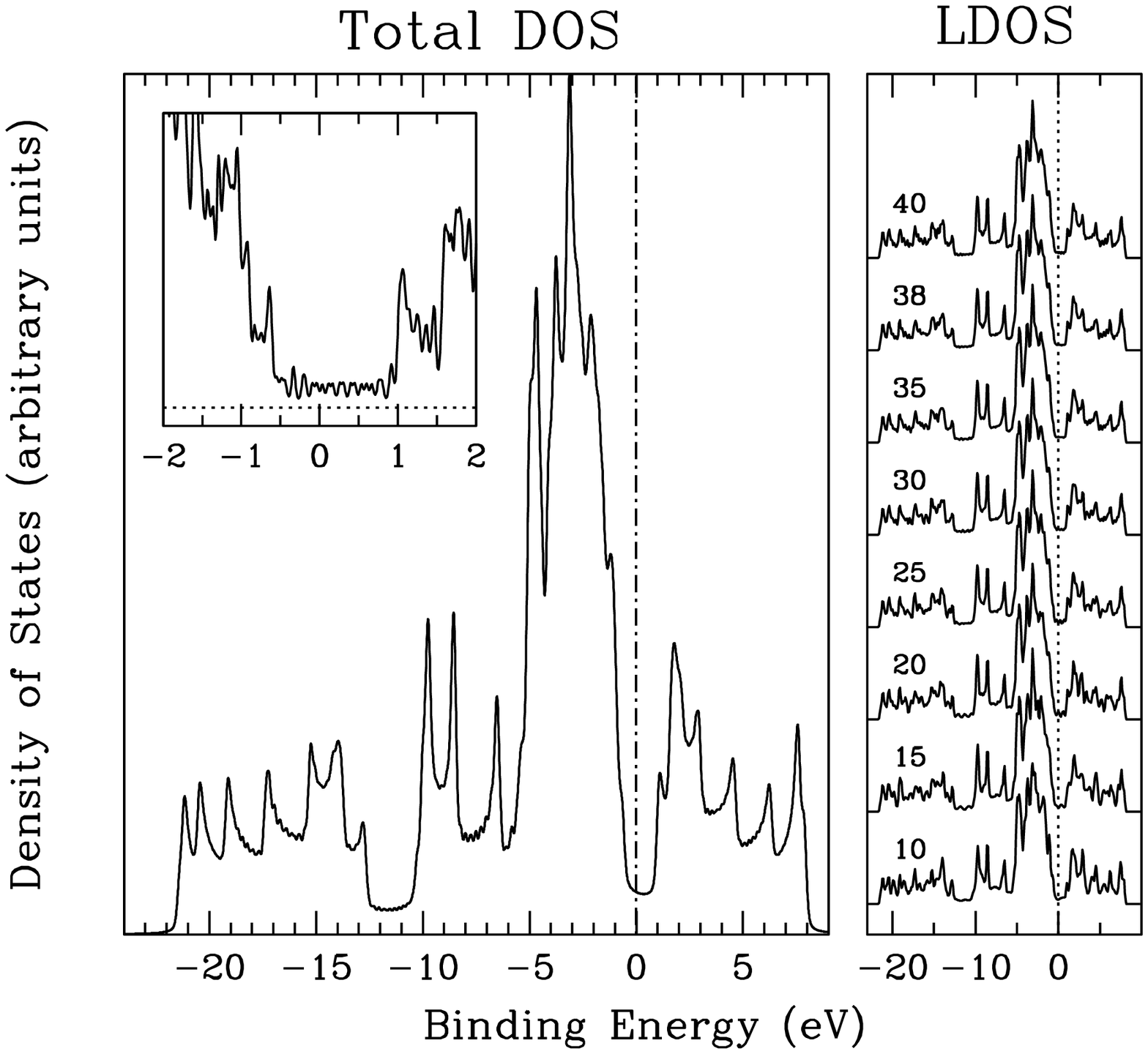} 
\end{center} 
~~\\
~~\\
Figure 1. Rochefort, Salahub and Avouris\\
\end{figure}
\thispagestyle{empty}

\clearpage
\begin{figure}[p] 
\begin{center}
\includegraphics[width=8cm,angle=-90]{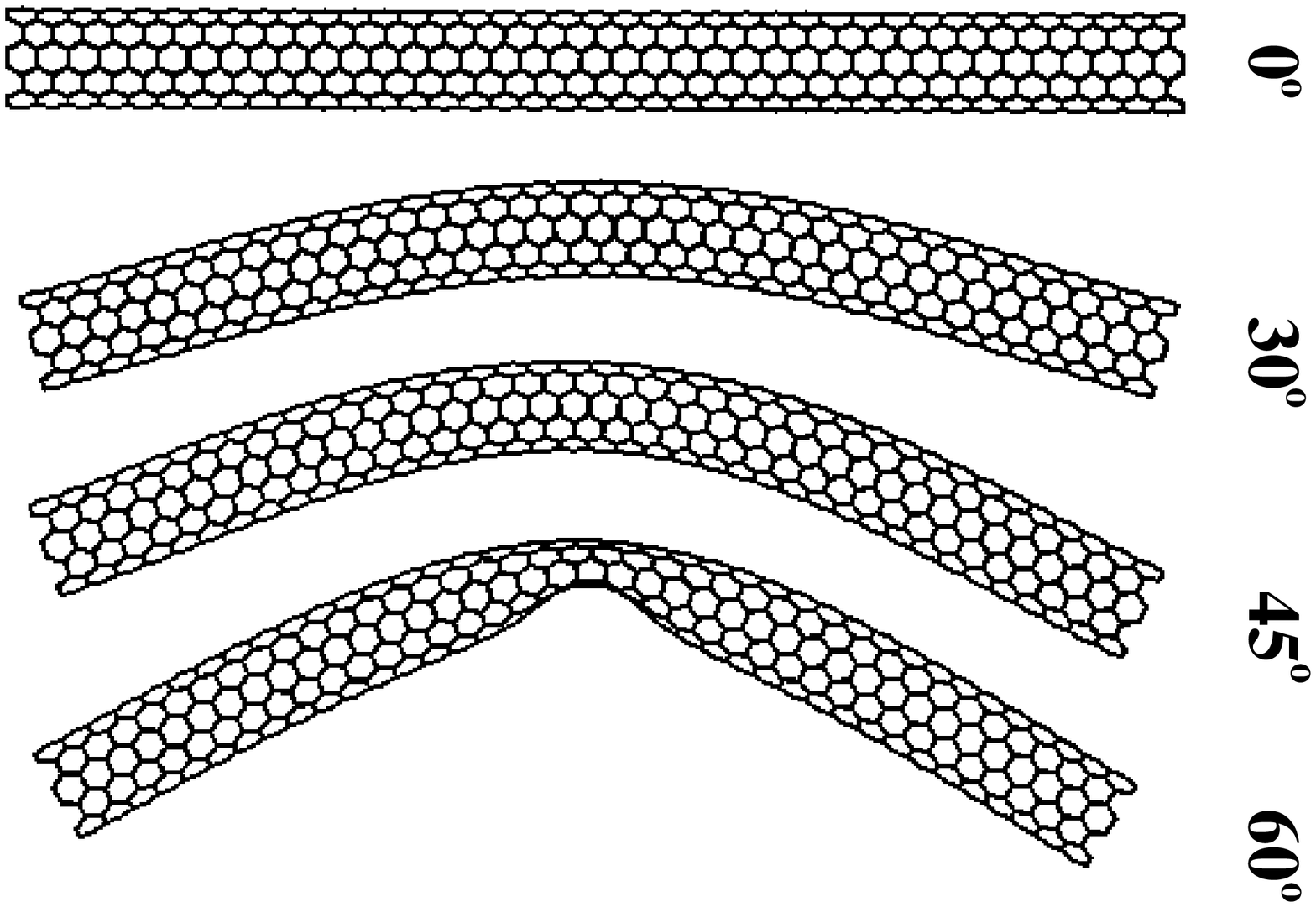}
\vskip0.2cm
\includegraphics[width=7cm,angle=-90]{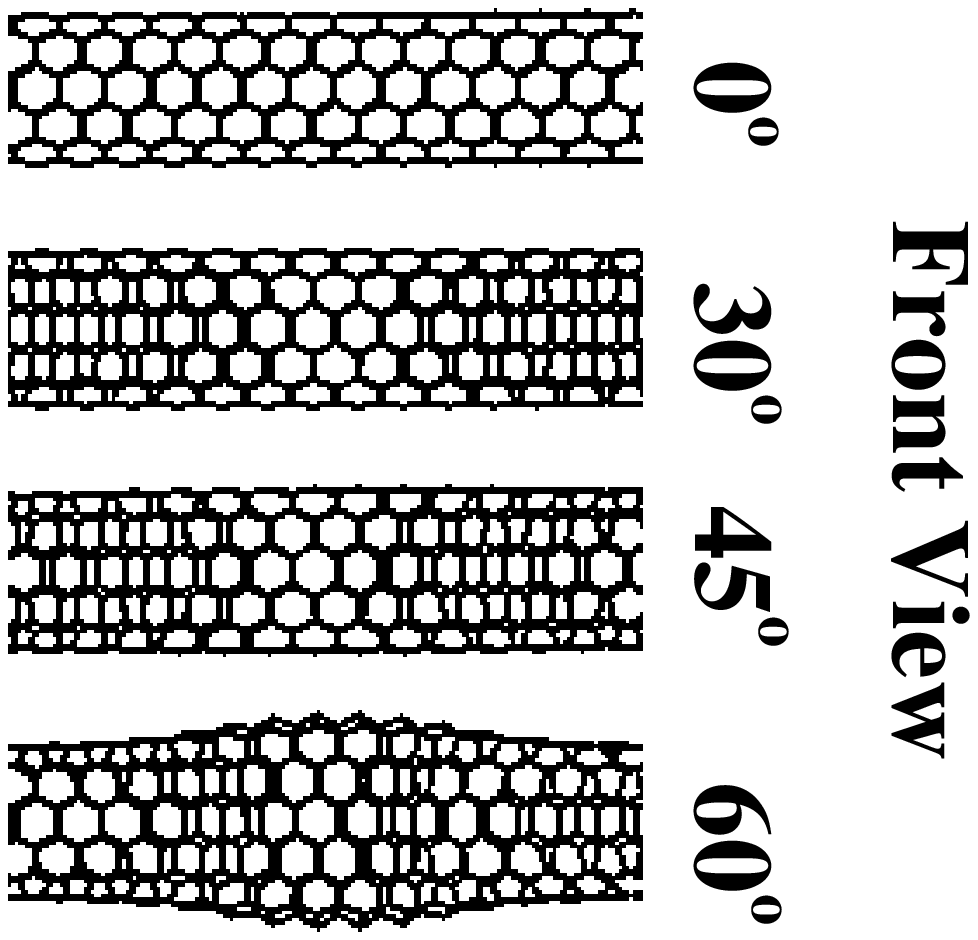}
\end{center} 
~~\\
~~\\
Figure 2. Rochefort, Salahub and Avouris\\
\end{figure}
\thispagestyle{empty}

\clearpage
\begin{figure}[p] 
\begin{center}
\includegraphics[width=15cm]{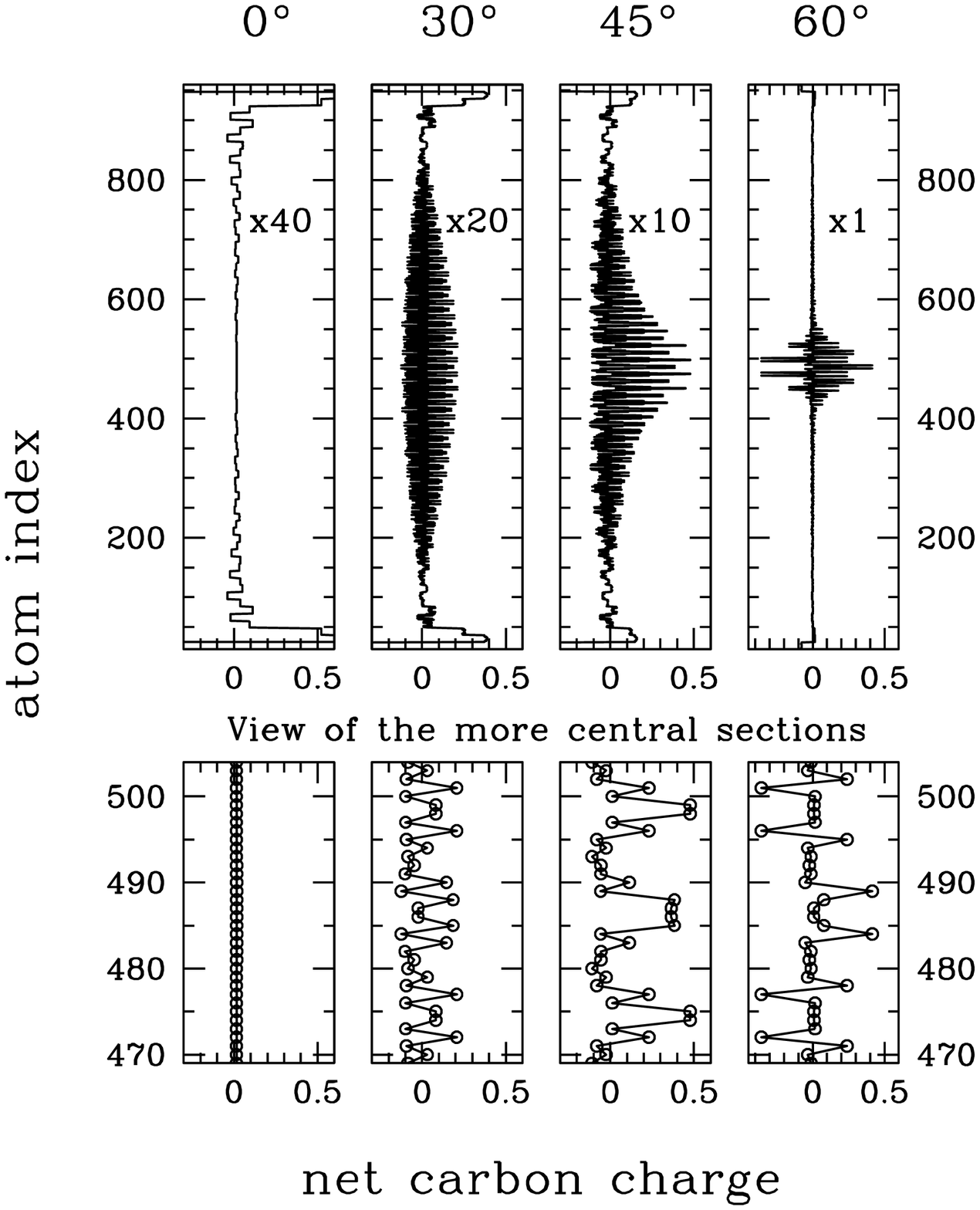} 
\end{center} 
~~\\
Figure 3. Rochefort, Salahub and Avouris\\
\end{figure}
\thispagestyle{empty}

\clearpage
\begin{figure}[p] 
\begin{center}
\includegraphics[width=15cm]{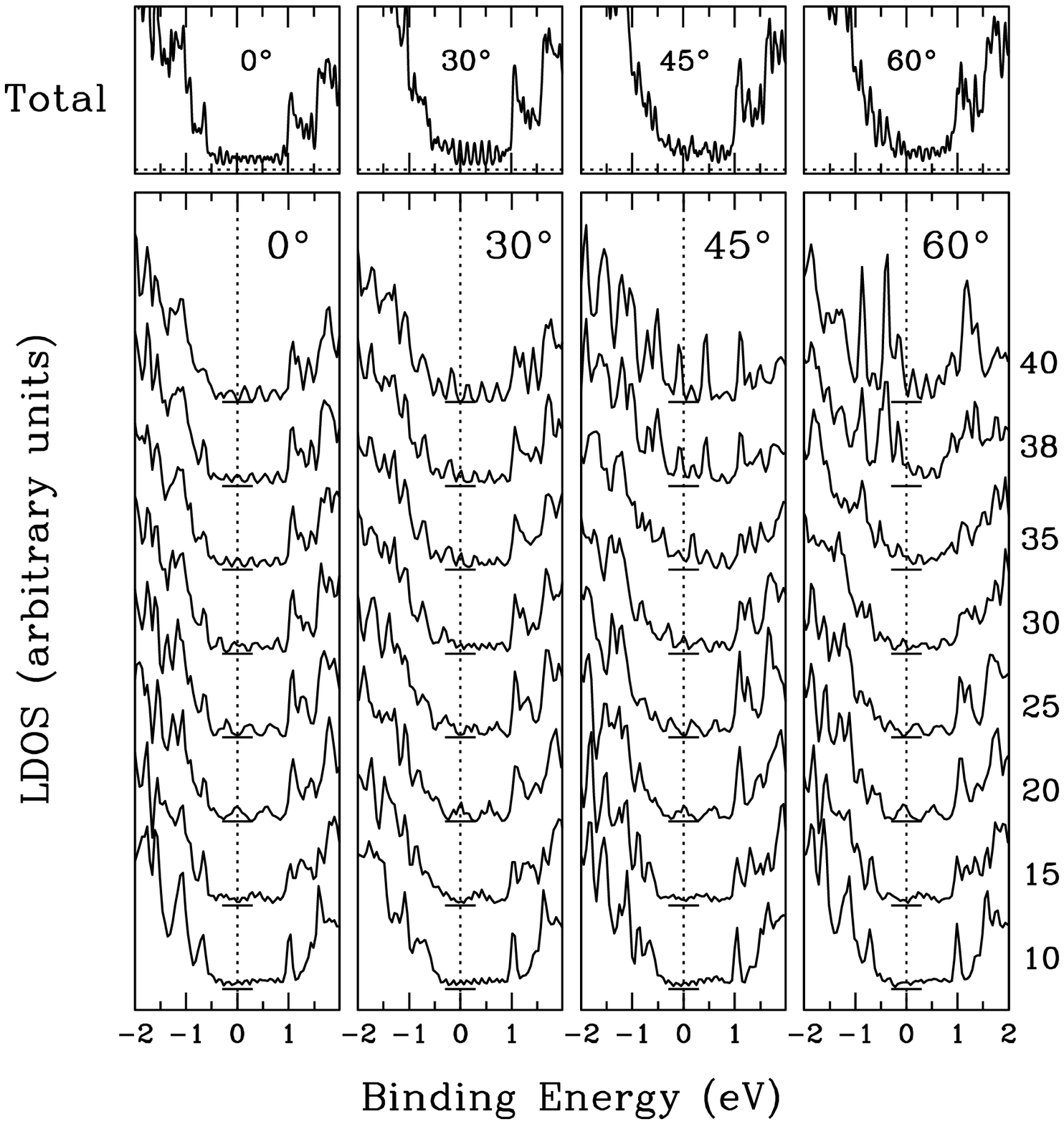} 
\end{center} 
~~\\
~~\\
Figure 4. Rochefort, Salahub and Avouris\\
\end{figure}
\thispagestyle{empty}

\end{document}